\documentclass[aps,pra,twocolumn,superscriptaddress,notitlepage]{revtex4-2}
\usepackage{CJK}
\usepackage[pdftex]{graphicx}
\usepackage{float}
\usepackage{amsmath}
\usepackage{amssymb}
\usepackage{url}
\usepackage{natbib}
 
\usepackage[colorlinks,
                   linkcolor=blue,
                   anchorcolor=blue,
                   citecolor=blue,
                   urlcolor=blue]
                  {hyperref}              
\begin{document}
\begin{CJK*}{UTF8}{gbsn}
\title{Response dynamics of alkali metal-noble gas hybrid trispin system}
\author{Guobin Liu}
\email{liuguobin@ntsc.ac.cn}
\affiliation{National Time Service Center, Chinese Academy of Sciences, Xi'an, 710600, China}
\affiliation{University of Chinese Academy of Sciences, Beijing, 100049, China}
\author{Vera Guarrera}
\affiliation{School of Physics and Astronomy, University of Birmingham, Edgbaston,
Birmingham B15 2TT, United Kingdom}
\author{Sihong Gu}
\affiliation{Innovation Academy for Precision Measurement Science and Technology, Chinese Academy of Sciences, Wuhan, 430071, China}
\affiliation{University of Chinese Academy of Sciences, Beijing, 100049, China}
\author{Shougang Zhang}
\affiliation{National Time Service Center, Chinese Academy of Sciences, Xi'an, 710600, China}
\affiliation{University of Chinese Academy of Sciences, Beijing, 100049, China}
\date{\today}
\begin{abstract}
We study the dynamics of a comagnetometer based on an alkali-metal/noble gas hybrid tri-spin system by numerically solving coupled Bloch equations. A thorough analysis of the response dynamics is carried out in the standard experimental regime, where the spin system can be mapped to a damped harmonic oscillator. The results show that a linear increasing response of the comagnetometer signal is found when the noble gas nuclear spin magnetization and the alkali spin lifetime parameters satisfy an over-damping condition. We find that an upper limit for the signal amplitude of the comagnetometer is imposed by the inherent dynamics of the hybrid tri-spin system. The results agree with currently available experimental data and provide useful guiding for future experiments. 
\end{abstract}
\maketitle
\end{CJK*}

The nuclear spins of noble gases can be easily polarized to a high degree by spin-exchange optical pumping \cite{Happer1997RMP}. Once polarized, nuclear spins can be hardly disturbed by the ambient static or electromagnetic fields due to their full shell electron configuration. These features make the nuclear spin polarization of noble gases a perfect resource for various scientific and practical purposes. As an example, trispin comagnetometers using one alkali metal and two noble gas species have recently become attractive for potential applications in Fundamental Physics measurement \cite{Bulatowicz2013PRL, Limes2018PRL}, and as a nuclear magnetic resonance (NMR) gyroscope for inertial navigation \cite{Donley2010NMRG, Walker2016NMRG} due to their exceptional sensing performances. 

The basic principle of these comagnetometers is to let the nuclear spin polarizations precess around a static magnetic field and then read these precessions out by using an alkali metal gas, which occupies the same volume of the noble gases' spins, as an ultra-sensitive in-situ magnetometer. Therefore, there are three kinds of polarized spins with different parameters occupying the same space, like an organic molecule with multiple spins in conventional NMR spectroscopy. 

Besides the polarization preparation and the magnetometer readout process, this hybrid trispin dynamic system itself displays a rich physical scenario deserving comprehensive study. The alkali metal-noble gas dual spin system has already revealed interesting nonlinear dynamics near the zero magnetic field, as shown both experimentally and theoretically for the K-$^3$He system \cite{Kornack2002PRL}. However, to the best of our knowledge, a similar thorough analysis for trispin systems, like Rb-${^3}$He-$^{129}$Xe \cite{Limes2018PRL} and Rb-$^{129}$Xe-$^{131}$Xe \cite{Korver2015PRL}, is still not available in experiment nor theory.

Recently we have studied the trispin dynamics by using numerical simulations of the coupled Bloch equations \cite{Liu2019PRA}. Empirical formulas were derived for the multiple-peak structure of the NMR spectra in the partial coupling configuration of the trispin system. Here, we would like to further extend the study of the trispin dynamics to disclose more information crucial for the operation of ultra-sensitive trispin comagnetometers.   

We assume a classical macroscopic magnetization field $\bf M$ is created for both the alkali vapour and the noble gases by mixed spin-exchange/optical pumping. Two basic interaction processes dominate the spin magnetization dynamics of the trispin system: the spin-field interaction between the atomic spin magnetizations and the static bias magnetic field ${\bf B}_0$, and the spin-spin interaction between the alkali spin magnetization and the spin magnetizations of the noble gases. 

The first interaction is well described by classical Bloch equations. Concerning spin-spin interactions, they have been shown to shift both the frequency of electron-paramagnetic-resonance (EPR) lines of alkali-metal atoms, and the NMR lines of noble-gas atoms \cite{Schaefer1989PRA}. These frequency shifts can be represented by introducing an effective spin-exchange field $\bf B_{se}$, which can be expressed as
\begin{equation}
{\bf B_{se}}=\lambda {\bf M}=\frac{8 \pi}{3}\kappa_0 \bf M.
\label{bse}
\end{equation}

This spin-exchange field is enhanced by a coefficient $\lambda$ in comparison to the classical macroscopic magnetization, due to the attractive force between the alkali atomic valence electron and the noble gas nucleus. The enhancement factor $\kappa_0$ is determined by the degree of overlap between the valence electron of the alkali atom and the noble gas nucleus \cite{Schaefer1989PRA}. It also depends weakly on the rate of spin-exchange collisions, and thus on the temperature of the gases \cite{Romalis1998PRA}. 

Following the approach used in \cite{Kornack2002PRL}, the linear coupled Bloch equations of the trispin system read: 
\begin{equation}
\small
\begin{split}
\dot{\bf M}^e&=\frac{\gamma_e}{q}{{\bf M}^e}\times[{\bf B}_0+\lambda_1 {{\bf M}^{n_1}}+\lambda_2{{\bf M}^{n_2}}] +\frac {M_0^e{\hat z}-{\bf M}^e} {\it{q T^e}},\\
\dot{\bf M}^{n_1}&=\gamma_1{{\bf M}^{n_1}}\times[{\bf B}_0+\lambda_1{\bf M}^{e}+\lambda_3{\bf M}^{n_2}]+\frac {{\it M}_0^{n_1}{\hat z}-{\bf M}^{n_1}} {[{\it T}_2^{n_1}, {\it T}_2^{n_1}, {\it T}_1^{n_1}]},\\
\dot{\bf M}^{n_2}&=\gamma_2{{\bf M}^{n_2}}\times[{\bf B}_0+\lambda_2{\bf M}^{e}+\lambda_3 {{\bf M}^{n_1}}]+\frac{M_0^{n_2}{\hat z}-{{\bf M}^{n_2}}} {[{\it T}_2^{n_2}, {\it T}_2^{n_2},{\it T}_1^{n_2}]}.
\label{eq:cbe1}
\end{split}
\end{equation}

Here, the $e$ in the subscript and superscript stands for the collective alkali atomic spin, which is mainly contributed by the outmost valence electron spin. $n_1$ and $n_2$ refer to the nuclear spins of the noble gas species 1 and 2 respectively. $T_1$ and $T_2$ are the longitudinal and transverse spin relaxation times for the corresponding spin species, as indicated by the superscript. $M_0$ is the initial spin magnetization, and $q$ is the slowing-down factor due to nuclear angular momentum \cite{Allred2002PRL}. Regarding the choice of the initial conditions, the nuclear spin magnetization has to be orthogonal to the direction of the bias magnetic field (${\bf B}_0$, along the $z$ axis) in order to acquire a nonzero torque from the bias field. Without loss of generality and for convenience, we let alkali atomic spin magnetization lie along the $x$ direction and nuclear spin magnetization along the $y$ direction.

Previously, we have theoretically shown that alkali metal-noble gas spin decoupling is necessary to realize precision comagnetometer without frequency shifts \cite{Liu2019PRA}. In practice, this can be done by modulating the alkali spins with an electro-optical modulator (EOM), by periodic optical pumping \cite{Korver2015PRL}, or by using NMR pulse decoupling techniques \cite{Limes2018PRL}. 
Indeed, in the presence of fast periodic modulations of the alkali spins, their precession around the dc bias field and the precession of the nuclear spins around the alkali spin magnetization, are both effectively averaged out, leaving the alkali macroscopic spin sensitive only to the nuclear spin magnetization. In this case, the Bloch equations take the form
\begin{equation}
\small
\begin{split}
\dot{\bf M}^{e}&=\frac{\gamma_{e}}{q}{{\bf M}^{e}}\times[\lambda_1 {{\bf M}^{n_1}}+\lambda_2{{\bf M}^{n_2}}] +\frac {M_0^{e}{\hat z}-{\bf M}^{e}} {\it{q T^{e}}},\\
\dot{\bf M}^{n_1}&=\gamma_1{{\bf M}^{n_1}}\times[{\bf B}_0+\lambda_3{\bf M}^{n_2}]+\frac {{\it M}_0^{n_1}{\hat z}-{\bf M}^{n_1}} {[{\it T}_2^{n_1}, {\it T}_2^{n_1}, {\it T}_1^{n_1}]},\\
\dot{\bf M}^{n_2}&=\gamma_2{{\bf M}^{n_2}}\times[{\bf B}_0+\lambda_3 {{\bf M}^{n_1}}]+\frac{M_0^{n_2}{\hat z}-{{\bf M}^{n_2}}} {[{\it T}_2^{n_2}, {\it T}_2^{n_2},{\it T}_1^{n_2}]}.
\label{eq:cbe2}
\end{split}
\end{equation}

At this point we can map the trispin system (and its dynamics) to a harmonic oscillator, based on the following experimental facts: 

1) once polarized and without significant interference, the noble gas nuclear spins can keep their polarization for tens of seconds or even longer (e.g. hours for $^3$He spin). In comparison, the alkali atomic spin polarization is short lived, typically decaying in 1 to 100 ms, and very sensitive to the ambient conditions, such as temperature and buffer gas pressure. In most cases the condition $T^n\ge10^4 T^e$ holds; 

2) with a heavier particle mass, the nuclear spin carries a smaller magnetic moment than the electron spin, determining a weaker spin-field interaction with the magnetic field and a smaller gyromagnetic ratio. Specifically for the case of interest for this work, the nuclear spin gyromagnetic ratio is typically about $10^3$ times smaller than that of the electron spin, i.e. $\sim$1 MHz/G versus $\sim$1 kHz/G. 

Therefore, we can consider the nuclear spin motion as a sinusoidally varying driving force and the alkali atomic spin motion as a damped harmonic oscillator. While the damping effect of the nuclear harmonic oscillators can be neglected due to their long spin lifetime, the alkali atomic spin oscillator has a significant damping rate, i.e. the inverse of its lifetime $\gamma$=2$\pi$/$q T^e$. The natural frequency of the alkali atomic spin oscillator is the Larmor frequency, i.e. $\omega_0$=$\gamma_e \lambda M^n /q$, which depends on the amplitude of the nuclear spin magnetization $M^n$. In this picture, the critical point between under and over damping of the alkali atomic spin oscillation is found for $\omega_0$=$\gamma$, which gives the condition: 
\begin{equation}
\lambda {\bf M}^n\it T^e=\rm 2\pi/\gamma_e.
\label{eq:cc}
\end{equation}

\begin{figure}[H]
\centering
\includegraphics[width=0.52\textwidth]{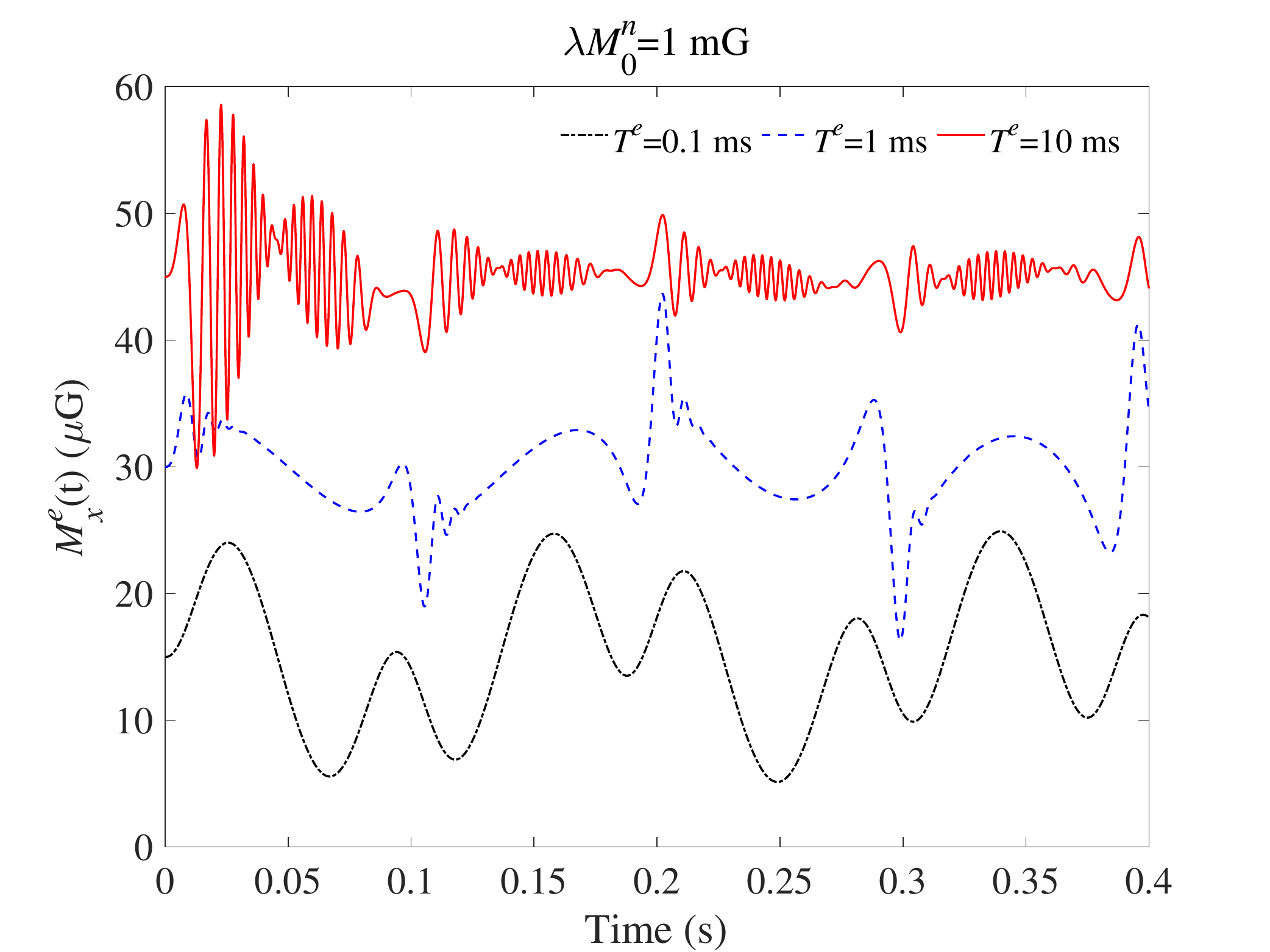}
\caption{(color online) Time evolution of the alkali atomic transverse spin magnetization $\it M_x^e$(t) calculated for different spin lifetimes $\it T^e$. The electron spin oscillation is seen as a damped harmonic oscillator, with critical damping condition determined by $\lambda \bf M^n \it T^e$=2$\pi$/$\gamma_e$. As $\it T^e$ increases, the alkali atomic spin oscillation moves from over damping to under damping, corresponding to an adiabatic and non-adiabatic following of alkali atomic spin to noble gas nuclear spin, respectively. The $T_1$ and $T_2$ spin relaxation times are taken 1\,hour and 1000\,s for $^3$He, 1000\,s and 80\,s for $^{129}$Xe. The enhancement factor $\kappa_0$ is taken as 5, 500 and -0.011 for Rb-He \cite{Romalis1998PRA}, Rb-Xe \cite{Ma2011PRL} and He-Xe \cite{Limes2019PRA, Vaara2019PRA} spin-exchange interactions, respectively.}
\label{fig1}
\end{figure}   

As a demonstration of this critical damping mechanism, we show the results obtained by numerically solving Eqs.\,\ref{eq:cbe2}. Fig.\,\ref{fig1} shows the initial 0.4 seconds of oscillation of the transverse alkali atomic spin magnetization for different values of the spin lifetime. These simulations have been done for $^{87}$Rb with $\gamma_e$=2$\pi \times$0.7 MHz/G, with critical point $\lambda \bf M^n \it T^e$=2$\pi$/$\gamma_e \approx$1.43 mG$\cdot$ms. 

In Fig.\,\ref{fig1}, it can be seen that for $\lambda M^n_0 T^e$=10 mG$\cdot$ms$>$1.43 mG$\cdot$ms, e.g. with $T^e=$10\,ms, the alkali atomic spin oscillation is under damped. Indeed one can observe a fast oscillation at the damped frequency $\omega_1$=$\sqrt {\omega_0^2-\gamma^2}$. Note that, as ${\bf M}^n$ is a sinusoidal function of time, $\omega_1$ is also dynamically changing. For $\lambda M^n_0 T^e$=0.1 mG$\cdot$ms$<$1.43 mG$\cdot$ms, e.g. with $T^e=$0.1\,ms, the alkali atomic spin oscillation is instead over damped and it adiabatically and smoothly follows the slow precession of the nuclear spin oscillations. Finally, when $\lambda M^n_0 T^e\simeq $1 mG$\cdot$ms, with $T^e=$1\,ms, the alkali atomic spin magnetization oscillates at a much smaller frequency and is shortly damped to its quasi-steady-state, which is completely determined by the nuclear spin precession.

Even though the long coherence time of the noble gas nuclear spin has been longly referred to as an advantage for atomic comagnetometer and NMR gyroscopes, the underlying mechanism of the coherence transfer between the different species of spins is not trivial. Here, our analysis of the critical damping mechanism provides a convenient way of understanding it. We also note that compared to other coherence transfer mechanisms in different spin systems \cite{Haroche1970PRL, Konrat1991JACS}, this coherence transfer naturally occurs with properly set parameters of the hybrid spin system, which can be easily manipulated \cite{Katz2015PRL}.

It is interesting to note that in pure alkali spin magnetometry, longer atomic coherence lifetimes lead to better sensitivity for the magnetic field detection. However, in a hybrid trispin comagnetometer, a strong damping factor, which means a short alkali atomic spin lifetime, is needed to realize a good coherence transfer from the noble gas nuclear spin to the alkali atomic spin during their coupled evolution.

More precisely we have found that, for a properly working comagnetometer, it is required that both the nuclear spin polarization and the alkali atomic spin lifetime are relatively small, so that their product remains below a critical value which depends on the specific magnetic moment of the alkali spin in use. However, for the merit of the comagnetometer signal, a strong nuclear spin magnetization is needed. This intrinsic conflict ultimately sets the limit of the comagnetometry response signal for a given trispin system, as we will see below.

By analyzing the case of under damping more in detail, we can see that the presence of high frequency oscillations prevents the alkali atomic spin motion from smoothly and adiabatically following the nuclear spin precession. In particular, if we decompose the overall evolution of alkali spins into a fast oscillation and an adiabatic slow oscillation, the fact that the fast oscillation and the adiabatic slow motion are not in phase leads to two main effects:

1) the phase of the overall oscillation is effectively advanced or retarded in a continuous way; 

2) the amplitude of the overall oscillation is reduced by different amounts, depending on the local phase relation. 

\begin{figure}[H]
\centering
\includegraphics[width=0.52\textwidth]{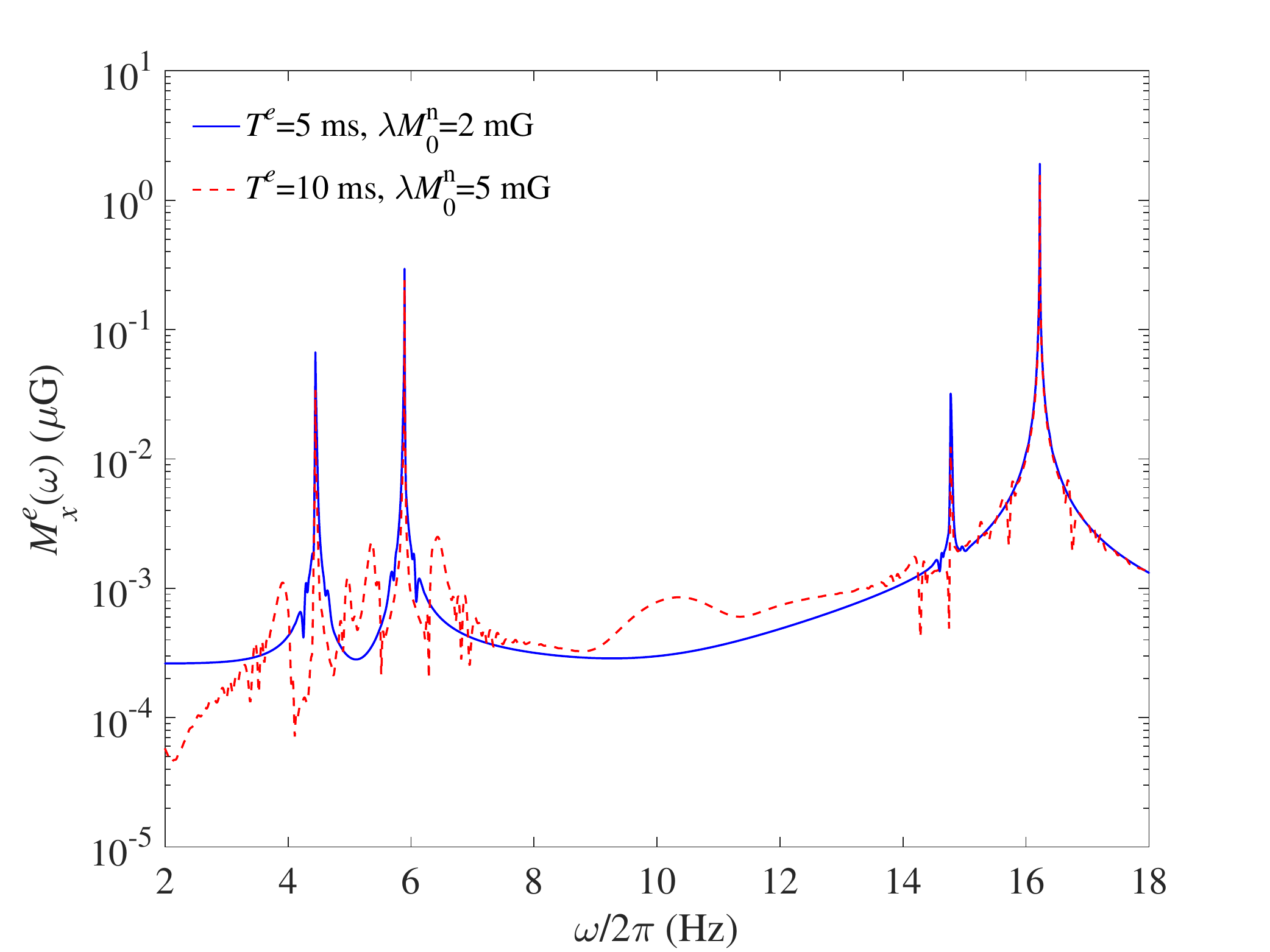}
\caption{(color online) Fourier transformation spectra for simulated 200 seconds oscillation of alkali atomic spin transverse magnetization. The spectra is shown as a function of initial nuclear spin magnetization $M_0^n$ and alkali atomic spin lifetime $T^e$. Extra side modes emerge near the main peaks as $M_0^n$ and $T^e$ increase, proving the gradual phase and frequency shift effects as a result of the under damping alkali atomic spin oscillations. The main resonance peaks are located at 5.89 Hz and 16.22 Hz, determined by the $^{129}$Xe and $^3$He atomic gyromagnetic ratios ($\gamma_{\rm Xe}$=2$\pi \times$1.178 kHz/G, $\gamma_{\rm He}$=2$\pi \times$3.244 kHz/G) and bias magnetic field $B_0$=5 mG.}
\label{fig2}
\end{figure}

The first effect leads to a continuous phase accumulation, and eventually generates a frequency shift. This is shown in the Fourier spectral analysis of Fig\,\ref{fig2} by the side modes appearing around the nuclear spin Larmor frequency (and associated peaks, as explained in \cite{Liu2019PRA}). The second effect leads to a reduction of the oscillation amplitude at the nuclear spin Larmor frequencies as $\lambda M^n_0 \it T^e$ increases (two highest peaks around 6 Hz and 16 Hz of dashed red line vs solid blue line in Fig\,\ref{fig2}). As a combined result, the NMR peaks at the nuclear spin Larmor frequencies become broader and lower in amplitude. This makes the overall spectral resolution worse, which, in turn, degrades the comagnetometer performances.

In a classical driven harmonic oscillator, with a small damping, a resonance amplification is possible when the natural frequency approaches the driving frequency. This also happens in the trispin system, however, in this case the over damping condition cannot be realized, and the NMR spectra become crowded with side modes components (similarly to the dashed red line in Fig\,\ref{fig2}). Ultimately, the original $^3$He and $^{129}$Xe resonance peaks, shown in our simulations, will be fully immersed in noise and not distinguishable any more. The system, in that case, becomes clearly unsuitable for high precision comagnetometer or gyroscope applications, and we will not consider this scenario any further. \color{black}

For practical purposes, we now discuss the maximum attainable comagnetometer response. Firstly, as we have explained above, this should be estimated in the linear
response range, i.e. in the over damping region where the alkali atomic spin follows the nuclear spin precession adiabatically. Secondly, the response is proportional to both the nuclear spin magnetization and the alkali atomic spin lifetime so that they shall be as high as possible. These two conditions determine the maximal comagnetometer response. 

\begin{figure}[H]
\centering
\includegraphics[width=0.52\textwidth]{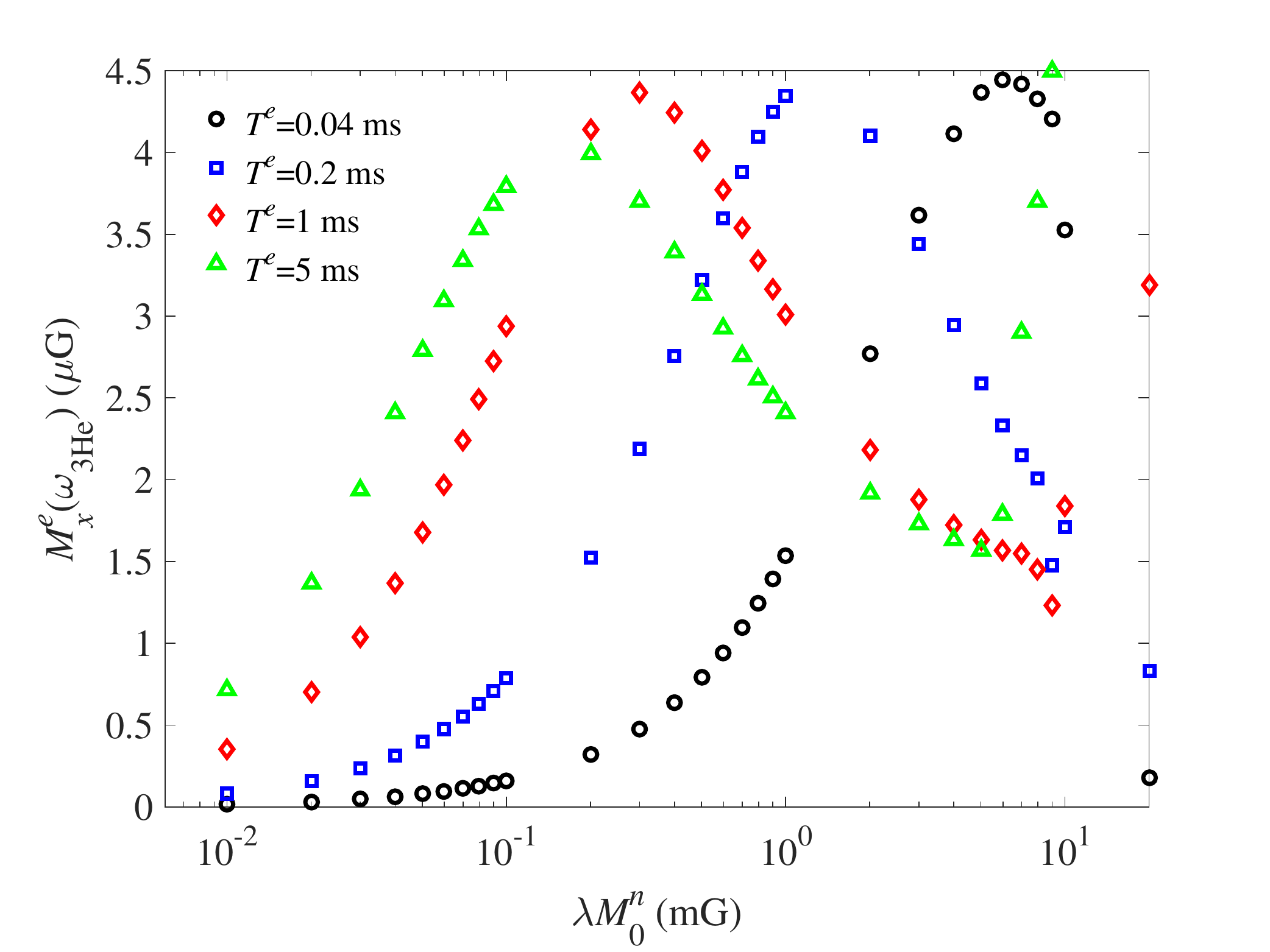}
\caption{(color online) The comagnetometer response $\it M_x^e$ at $^3$He spin precession Larmor frequency as a function of nuclear spin magnetization $\it M^n$ at various alkali atomic spin lifetimes $\it T^e$. A maximum response can be obtained over a large range, almost two orders of magnitude for both $\it M^n$ and $T^e$ parameters. The abnormal increasing response of the last few points (2 red squares and 4 green triangles) at high $\it M^n$ for $T^e$=1$\&$5 ms are region where the resonance amplification happens for small damping and strong driving.}
\label{fig3}
\end{figure}

To find out this maximum response, we simulated the alkali atomic spin evolution for 200 seconds in the time domain and then we did a Fourier transform to extract the peak value at the $^3$He Larmor frequency. We repeated the process for different values of the nuclear spin magnetization and alkali atomic spin lifetime. Finally, we plotted the comagnetometry signal amplitude as a function of the $\lambda M^n_0$ spanning a range of three orders of magnitude at different $T^e$, as shown in Fig.\,\ref{fig3}. 

A single maximum response signal is found as a function of the nuclear spin magnetization for different alkali atomic spin lifetimes. One can also see that when the alkali atomic spin lifetime is short, the response signal reaches the maximum at high nuclear spin magnetization, and vice versa. These results are in agreement with our previous analysis. The maximum response is almost constant over a range of two orders of magnitude for nuclear magnetization, from 0.1 mG to 10 mG and over a range of two orders of magnitude for alkali atomic spin lifetime, from 0.04 ms to 5 ms. This means that the system has a relatively large linear response regime. Fortunately, these ranges can be easily covered experimentally by tuning parameters such as the temperature and the buffer gas pressure. It's also interesting to notice that the best current experimental values, reported e.g. in \cite{Korver2015PRL, Limes2018PRL}, fall within the best region of parameter highlighted by our simulations. 

In addition, the response signal is a cross product of the alkali atomic spin magnetization ${\bf M}^e$ and noble gas nuclear spin magnetization ${\bf M}^n$, which means it is also proportional to the alkali atomic spin magnetization. Given a certain initial alkali spin magnetization $M^e_0$, the existence of the maximum response signal indicates that there is an upper limit for the comagnetometer output based on the alkali metal-noble gas hybrid spin system. Considering the critical damping condition of Eq.\,\ref{eq:cc}, the maximum response of the trispin comagnetometer is determined only by the alkali atomic spin parameters, i.e. by the spin magnetization and lifetime of the alkali atom gas.

Finally, we point out that the trispin system dynamics converges even in the presence of a multiple peak spectral structure, observed for some specific conditions in \cite{Liu2019PRA} and \cite{Limes2018PRL}. Even if a chaotic relaxation might not be excluded in these cases (see for example \cite{chaos2008, chaos2012} and references therein), from our numerical analysis this appears not to have a major impact on the significance of the critical damping condition, and on the main results outlined in the present work.

In conclusion, we have investigated the intrinsic response dynamics of a comagnetometer based on a gaseous alkali metal-noble gas hybrid trispin system. From the perspective of a damped harmonic oscillator, we revealed the important relations between the key atomic spin parameters. The alkali atomic spin oscillator has to work at the over damping condition, i.e. a relative low noble gas nuclear spin magnetization and short alkali atomic spin lifetime, to have a linear response and achieve the maximum signal. The over damping can also serve as a possible mechanism for coherence transfer between the alkali atomic spin and noble gas nuclear spins. 

For the benefit of a high performance comagnetometer or NMR gyroscope, counterintuitively, a compromise between strong noble gas nuclear spin magnetization and long alkali atomic spin lifetime shall be made. For future experiments on trispin NMR gyroscopes, a maximum achievable response is given as a function of the noble gas nuclear spin magnetization and alkali atomic spin lifetime. The limit is inherent in the response dynamics of the hybrid trispin system. We stress that, even though we used the $^{87}$Rb-$^3$He-$^{129}$Xe system for our simulations, the theoretical results obtained here are valid for any alkali metal-noble gas hybrid trispin system.   

We thank the financial support under the grant NO. E024DK1S01 by National Time Service Center, Chinese Academy of Sciences. We thank Joseph Nicholson for careful reading of the manuscript.

\end{document}